\documentclass[iop]{emulateapj}

\shortauthors{Ben-Jaffel et al.}
\shorttitle{The interstellar bow shock}
\newcommand{\lya}{Lyman-$\alpha$} 
\usepackage{natbib}

\begin{document}

\title{The existence and nature of the interstellar bow shock}

\author{Lotfi Ben-Jaffel \altaffilmark{1}}
\affil{UPMC Univ Paris 06, UMR7095, Institut d'Astrophysique de Paris, 
F-75014, Paris, France; bjaffel@iap.fr}
\altaffiltext{1}{CNRS, UMR7095, Institut d'Astrophysique de Paris, F-75014, Paris, France}

\author{M. Strumik, R. Ratkiewicz\altaffilmark{2}, J. Grygorczuk}
\affil{Space Research Centre, Polish Academy of Sciences, Bartycka 18A, 00-716 Warsaw, Poland}
\altaffiltext{2}{Institute of Aviation, Warsaw, Poland}

\begin{abstract}
{We report a new diagnostic between two different states of the local interstellar medium (LISM) near our solar system using a sensitivity study constrained by several distinct and complementary observations of the LISM, solar wind, and inner heliosphere. Assuming the Interstellar Boundary Explorer (IBEX) He flow parameters for the LISM, we obtain a strength of $\sim 2.7\pm0.2$\,$\mu$G and a direction pointing away from galactic coordinates $(28, 52)\pm 3^\circ$ for the interstellar magnetic field as resulting from fitting Voyager 1 \& 2 in situ plasma measurements and IBEX  energetic neutral atoms ribbon}. When using Ulysses parameters for the LISM He flow, we recently reported the same direction but a strength of $2.2\pm0.1$\,$\mu$G. First, we notice that with Ulysses He flow, our solution is in the expected hydrogen deflection plane (HDP). In contrast, for the IBEX He flow, the solution is $\sim 20^{\circ}$ away from the corresponding HDP plane. Second, the long-term monitoring of the interplanetary H\,I flow speed shows a value of $\sim 26$\,km/s measured at upwind from the Doppler-shift in the strong \lya\, sky background emission line. All elements of diagnostics seem therefore to support Ulysses He flow parameters for the interstellar state. In that frame, we argue that reliable discrimination between superfast, subfast, or superslow states of the interstellar flow should be based on most existing in situ  and remote observations used together with global modelling of the heliosphere. For commonly accepted LISM ionization rates, we show that a fast interstellar bow-shock should be standing-off upstream of the heliopause.
\end{abstract}

\received{June 19, 2013}
\accepted{October 20, 2013}


\section{Introduction} 
The detection of interstellar neutral hydrogen in the neighborhood of the Earth orbit in the seventies of the twentieth century opened a new era of investigation on the origin and properties of the magnetized LISM, particularly its interaction with the time-variable and anisotropic solar wind plasma and interplanetary magnetic field (IMF) \citep{ber71,tho71,axf72}. The idea of heliospheric distortion by the interstellar magnetic field (ISMF) was introduced since the late eighties using the very simple Newton approximation \citep{fah88}, a concept that was theoretically generalized later on using magnetohydrodynamic (MHD) 3D simulations that clearly predict an asymmetric heliosphere in the presence of an oblique ISMF \citep{rat98,pog98}. The first evidence for such an asymmetry was a $\sim 12^{\circ}$ deflection of the heliosphere's nose that was reported using backscattered \lya\, emission measured by the Voyager ultraviolet spectrometer deep in the heliosphere \citep{bpr00}. Using the \lya\, emission distrbution over the sky as remotely observed by the Solar Wind Anisotropies instrument onboard the Solar Wind Heliospheric Observatory (SoHO/SWAN), it was possible to find that the direction of the neutral hydrogen after crossing the heliopause is slightly deflected from the LISM He flow, another new indication on the heliosphere distortion \citep{lal05}. 

Meanwhile, Voyager 1 \& 2 (V1 \& V2) {\it in situ} measurements of the heliopsheric plasma and field reported along their trajectories out of the solar system revealed an asymmetric termination shock shape that is probably related to the distortion of the heliosphere \citep{bur05,dec05,gur05,sto05,bur08,dec08,gur08,ric08,sto08}. Nevertheless, the V1 \& V2 unique harvests of {\it in situ} measurements corresponding to a time evolving snapshot of the heliopsheric gas in two distinct directions of space are raising challenging questions that nowadays theoretical models cannot answer \citep{bur13,dec12,ric12,zan10}. Soon after the V2 termination shock (TS) crossing, a bright ribbon of energetic neutral atoms (ENAs) was imaged by IBEX \citep{mcc09}. We remark that Cassini observed high energy ENAs with a similar but less clear structure \citep{krim09}. The ribbon structure is probably governed by the ISMF interaction with the heliosphere \citep{sch09,heer10,chal10,sch13}.  

Recently it has been shown that the asymmetries observed in the sky background \lya\ emission deep in the heliosphere, in the solar wind (SW) TS position as measured by V1 and V2, and in the IBEX ribbon of ENAs, can all be explained with a rather weak $ |\mathbf{B}_{\mathbf{\infty}}|\sim 2.2\pm 0.1\, \mu$G ISMF pointing away from galactic coordinates $(28, 52)\pm 3^\circ $ \citep{br12}. This direction makes an angle of $\sim 42^\circ$ with Ulysses helium flow (see Table 1) in agreement with our first finding \citep{bpr00}. The ISMF vector derived there lies in the HDP \citep{lal10}, consistent with an independent analysis of the hydrogen flow deflection angle as observed by SOHO/SWAN \citep{heer11}. We have also shown that the MHD modeling of the only V1 and V2 {\it in situ} measurements or IBEX data set does not lead to unique values of the ISMF strength and orientation as previously expected. The asymmetry thus far reported for the TS shape is not only the result of the ISMF effect but also of the SW anisotropy and time variation \citep{br12}.

\begin{table}
\caption{LISM velocity and temperature: Ulysses- and IBEX-based He parameters. Inflow direction in ecliptic coordinates ($\lambda$, $\beta$).}
\begin{center}
\hspace*{-0.1in}
\scalebox{0.85}{
\begin{tabular}{cccccc}
\hline
He parameters & $\lambda$ & $\beta$ & \mbox{$V_{\mathrm{\infty}}$} & \mbox{$T_{\mathrm{\infty}}$} & Ref. \\
& & & & & \\
Ulysses & $255.4^\circ$ & $5.2^\circ$ & $26.4$ km/s & $6400$ K & \citep{wit04} \\
IBEX & $259.2^\circ$ & $5.1^\circ$ & $22.8$ km/s & $6200$ K & \citep{bzo12} \\ 
\hline
\end{tabular} }
\end{center}
\end{table}

Nevetheless, a controversy is persisting about the existence of the interstellar bow shock, with opposite diagnostics originating from studies that are using different LISM parameters. For example, for the 2000 epoch Ulysses-based helium parameters, the interstellar bow shock is claimed to exist \citep{pog09,br12}. However, recent observations from the IBEX satellite show that the relative motion of the Sun with respect to the interstellar medium is apparently slower than the fast magnetosonic speed \citep{bzo12,mcco12}. This leads to a plasma configuration with no fast bow shock forming ahead of the heliosphere, in contrast to the idea  widely expected in the past. The problem is that such heliosphere configuration produces plasma and field distributions that are not consistent with the Voyagers TS crossings. Indeed, \citet{br12} showed a direct comparison between their MHD model and Voyager 1 \& 2 TS crossing observations when using the IBEX He direction and the field strength ($\sim 3\,\mu$G) derived by \citet{mcco12}. That comparison clearly shows a misfit for both TS crossings (see their Fig. 2 \& 4). We are not aware of any other publication in the literature showing such direct comparison when using the IBEX He parameters. The conclusion is that the IBEX He speed and the field strength derived by McComas et al. 2012 are not compatible with the Voyager TS crossings \citep{br12}. 

Admittedly \citet{zan13} considered IBEX LISM parameters jointly with TS crossings in their study of the general problem of formation of shocks mediated by neutrals. However they assumed the same TS location ($\sim 89$\,AU) for both the V1 and V2 crossings, which is significantly different from the exact V1 \& V2 TS crossings ($\sim$ 83.7 AU and $\sim$ 94 AU). On the other hand considering the subfast LISM interaction, a slow bow shock ahead of the heliopause can alternatively be produced \citep{pog06,zie13}. However, MHD theory allows a slow bow shock provided that the angle between the LISM velocity and magnetic field vectors is small \citep{zie13}. This strong constraint on the magnetic field orientation contradicts results obtained from models reproducing at least the shape of the measured IBEX ribbon \citep{sch09,gry11,heer11,stru11,br12,rat12,sch13}.

In this paper we present a continuation of our sensitivity study but use the LISM He flow parameters as reported by the IBEX team (see Table 1). Namely, we try to fit simultaneously the IBEX ribbon and both Voyagers crossings using MHD 3D simulations in which we test a large set of $\sim 2100$ models of different strengths and orientations of the ISMF, many SW activities for the period 2004-2009, and a fine 3D grid.  Finally, the proposed study tests whether available observations and simulation tools are adequate to discriminate between two different boundary conditions of the LISM in the neighborhood of our solar system.

\section{Model and method}
To study the SW-LISM interaction, we use the same set of MHD equations that have a source term {\bf S} on the right-hand-side (RHS) describing charge exchange with the assumed constant flux of hydrogen \citep{rb02}:

\begin{equation}
\frac{\partial {\bf U}}{\partial t} +
{\bf \nabla}\cdot{\bf \bar{F}} = {\bf Q + S }
\end{equation}

where ${\bf U}$,  ${\bf Q}$, and ${\bf S}$ are column vectors,
and ${\bf \bar{F}}$ is a flux tensor defined as:

$$
\begin{array}{ll}
{\bf U} = \left|            
\begin{array}{c} \rho \\ \rho {\bf u} \\ {\bf B} \\ {\mathrm{\rho E}} \\ 
\end{array} \right|
& \hspace{1.6cm}
{\bf \bar{F}} = \left|
\begin{array}{c} \rho {\bf u}\\
\rho {\bf u}{\bf u} + {\bf I}(p + \frac{{\bf B} \cdot {\bf B}}{8 \pi})
- \frac{{\bf B}{\bf B}}{4 \pi}\\
{\bf u}{\bf B} - {\bf B}{\bf u} \\
\mathrm{\rho H} {\bf u}-\frac {{\bf B}({\bf u} \cdot {\bf B}}{4 \pi})\\
\end{array} \right|
\end{array}
$$

$$
\begin{array}{ll}
{\bf Q} = - \left|
\begin{array}{c} 0 \\ \frac {\bf B}{4\pi} \\ {\bf u}
\\ {\bf u} \cdot \frac {\bf B}{4 \pi} \\
\end{array} \right| \nabla \cdot {\bf B} \end{array} $$

$$
\begin{array}{ll}
{\bf S} = \rho {\nu}_c \left|
\begin{array}{c} 0 \\  {\mathrm{{\bf V}_H - u}} \\ {\bf 0}
\\ \mathrm{ \frac {1}{2}{V_H}^2 + \frac {3k_BT_H}{2m_H} - \frac {1}{2} u^2
- \frac {k_BT}{(\gamma - 1) m_H}} \\
\end{array} \right|
\end{array}
$$

Here, $\mathrm{\rho}$ is the ion mass density, $\mathrm{p=2nk_BT}$ is the 
pressure, $\mathrm{n}$ is the ion number density, $\mathrm{T}$ and 
$\mathrm{T_H}$ ($\mathrm{T_H}={\mathrm const}$) are ion and H atom temperatures, and  $\mathrm{{\bf u}}$ and 
$\mathrm{{\bf V}{_H}}$ ($\mathrm{{\bf V}_H}={\mathrm \bf const}$) are the 
ion and H atom velocity vectors, respectively; $\mathrm{{\bf B}}$ is the
magnetic field vector, $\mathrm{E=\frac{1}{\gamma-1} \frac{p}{\rho} +
\frac{\bf u \cdot u}{2} + \frac{\bf B \cdot B}{8 \pi \rho}}$ is the
total energy per unit mass, and $\mathrm{H =
\frac{\gamma}{\gamma-1} \frac{p}{\rho} + \frac{\bf u \cdot u}{2} +
\frac{\bf B \cdot B}{4 \pi \rho}}$ is enthalpy. $\mathrm{\gamma}$ is
the ratio of specific heats. {\bf I} is the 3 x 3 identity matrix. 
The charge exchange collision frequency is $\mathrm{{\nu}_c = n_H \sigma u_*}$, where $\mathrm{n_H}$ is H atom number density,  
$\mathrm{\sigma}$ is the charge exchange cross-section, and $\mathrm{u_* = (({\bf u} - {\bf V}{\mathrm{_H}})^2 + 128 k_B(T + T_H)/ 
(9 \pi m_H))^{1/2}}$ is the effective average relative speed of protons and H atoms, assuming that both protons and H atoms velocities follow a maxwellian distribution. The flows are taken to be adiabatic with $\mathrm{\gamma=5/3}$. The divergence-free magnetic field constraint, $\mathrm{\nabla\cdot {\bf B}=0}$, is expressed in the numerical simulations incorporating the source term {\bf Q} to the RHS of (1), which is proportional to the divergence of the magnetic field. By adding {\bf Q} to the RHS of (1) assures that any numerically generated $\mathrm{\nabla \cdot {\bf B} \neq  0}$ is advected with the flow, and allows one to limit the growth of $\mathrm{\nabla \cdot {\bf B} \neq  0}$.

The three-dimensional set of equations defined in Eq. 1 is solved using a spatial first order time-marching, implicit, upwind-differenced scheme based on a finite-volume approach \citep{rat98}. The spherical grid used for the computations is generated internally in the 3D MHD code, with 100 grid points in the azimuthal, 146 points in the radial, and 200 points in the meridional direction.

In the present study, we used the MHD model described above with a magnetized interstellar plasma where the solar wind is also magnetized and its latitudinal anisotropy is included. The simulation frame is Sun-centered, the inflow LISM velocity vector $\mathbf{V}_{\mathbf{\infty}}$ defines the x-axis and the ISMF vector
$\mathbf{B}_{\mathbf{\infty}}$ lies in the x-y plane. The ISMF orientation is determined by the inclination angle $\alpha$ (between
$\mathbf{V}_{\mathbf{\infty}}$ and $\mathbf{B}_{\mathbf{\infty}}$ vectors) and the deviation angle $\beta$ between
$\mathbf{V}_{\mathbf{\infty}}$-$\mathbf{B}_{\mathbf{\infty}}$ plane and the HDP \citep{lal10}. We used a five-degree spaced regular grid for testing
different orientations of the ISMF in the range $0^{\circ} \leq \alpha \leq 90^{\circ}$ and $-90^{\circ} \leq \beta \leq 90^{\circ}$. The ISMF
strength values used in our study are $2.0$, $2.2$, $2.4$, $2.8$, $3.2$, and $3.6$\,$\mu$G. After an approximate solution is derived over the selected ISMF grid, a final MHD simulation is calculated with the obtained fit.

The flux of hydrogen was assumed to be constant \citep{rb02}, where the neutral H number density $n_\mathrm{H}$ = 0.11 cm$^{-3}$ is in the range $0.1 - 0.2$ cm$^{-3}$ indicated by \citet{izm03}. It is important to stress that the asymmetry of the TS is not affected by the treatment of neutrals, provided the density at the TS location is consistent with observations \citep{izm03,alo11}. Though using a simplified treatment of neutrals, our approach provides reliable estimations for ISMF parameters as recently shown in comparison to results obtained from independent models using a full kinetic description for neutrals \citep{heer11,stru11,br12}. 

In addition to the parameters shown in the second row of Table 1, the LISM plasma density $n_{p}=0.095$ cm$^{-3}$ is used in the outer boundary condition for our simulations. The selected LISM plasma density is in the middle of the density range thus far reported $n_{p} \sim 0.05-0.13\, {\rm cm^{-3}}$ in the literature \citep{sla08}, yet the impact of the full density range on the existence of the insterstellar bow-shock will be discussed in section 3.3.

\begin{figure}
\centering
\includegraphics[width=9.cm]{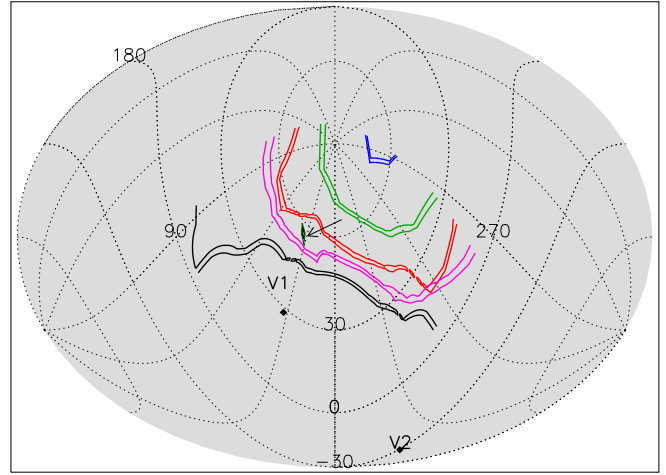}
\caption{Contours of regions in galactic coordinates of the ISMF orientations where the MHD model reproduces the TS crossing for V2 on 2007 August 30 for the ISMF strength equal to (blue) 2.0, (green) 2.2, (red) 2.4, (pink) 2.8, and (black) 3.6 $\mu$G (contours are shown from top to bottom in increasing strength order). For the IBEX ribbon, the small black contour (indicated by an arrow) shows $|\mathbf{B}\cdot \mathbf{R}|<0.03$ obtained for $2.7 \mu$G as the best fit (embedded between red (2.4) and orange (2.8) contours at latitudes higher than $30^{\circ}$). The $\mathbf{B}\cdot\mathbf{R}=0$ condition does not show a high sensitivity to the ISMF strength and only the best solution is shown.}
\end{figure}

{Ideally, time- and latitude- dependent solar wind boundary conditions should be included for every set of the LISM parameters listed above ($\sim 2100$ cases). At present, time dependent simulations of the heliosphere are feasible for fixed values of LISM parameters \citep{was11,pog13}. However, a comprehensive sensitivity study using fully time-dependent approach is beyond the presently available computational resources. Here, solar cycle effects are included in a simplified manner. We tested five values} of the slow solar wind speed in the range from 420 to 500 km/s, consistent with majority of measurements made by the plasma detector on V2 and by the {\it Advanced Composition Explorer} (ACE), an Earth-orbiting satellite. The fast wind speed is assumed $\sim 1.9$ times the slow speed value \citep{tok10}. We use yearly maps of observed interplanetary scintillations provided by \citep{tok10} to fix the latitudinal extent of the slow and fast solar wind in our model. These observations suggest that the fast wind cone extends outside the latitudes range $\sim \pm56\pm 6^\circ$ for the period of 2001-2005, and $\sim\pm36\pm 6^\circ$ for the period of 2006-2008 (heliographic latitudes measured from Sun equator). During the intermediate period of 2005-2006 the fast wind cone was assumed to shrink linearly between the two values. The plasma density at 1 AU is assumed $ n_\mathrm{SW}=5.2$\, ${\rm cm}^{-3}$, thus the corresponding SW dynamic pressure (in the range of $1.4-2.1 $\,nPa) is proportional to the square of the SW speed variation. The solar equator is tilted $7.25^{\circ}$ with respect to the ecliptic plane and the ecliptic longitude of the Sun's ascending node is $75.77^{\circ}$(J2000). The interplanetary magnetic field has the form of Parker's spirals, where the radial component is equal to 35.5\,$\mu$G at 1 AU.

\begin{figure}
\centering
\hspace*{-0.4in}
\includegraphics[width=9.5cm]{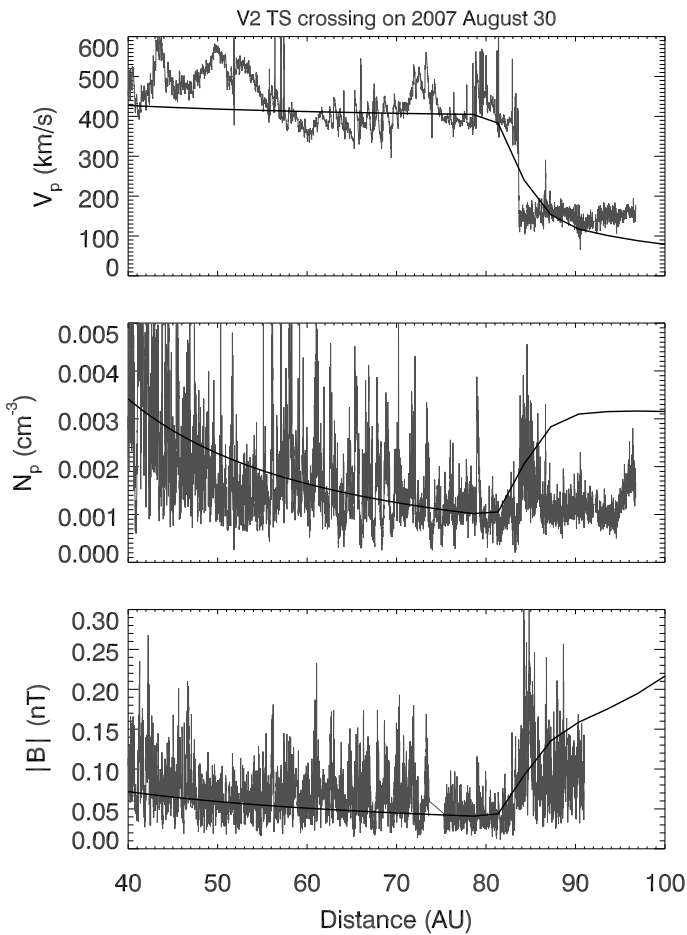}
\caption{ Comparison of the V2 measurements (gray) and MHD model results (black solid) 
for simulation parameters $\alpha=45^\circ$, $\beta=20^\circ$, ISMF strength $2.7\, \mu$G, and a SW speed of $\sim 420$km/s. This model is the best solution obtained when using the IBEX He parameters (Fig. 1). From the top panel, we show the plasma speed, density, and magnetic field strength.}
\end{figure}

Assuming the IBEX-based He parameters, the estimation of the ISMF parameters consists in finding a simulation case that gives the best fit to existing observations. In the first step, we test the model simultaneously against the IBEX ribbon observations and the TS crossing by V2 using {\it in situ} plasma and magnetic field measurements. The solar wind speed and density are not available from plasma experiment on V1, but we assume that they can be derived from V2 plasma measurements obtained at the time of TS crossing by V1 and ACE observations obtained about one year earlier and propagated properly from the ACE location to the V1 position. This procedure allows us to test the model also for TS crossing by V1. This kind of approach was shown to work well in our previous studies \citep{stru11,br12}, thus should also provide an ISMF vector consistent with all observations but obtained for the new IBEX He flow parameters.

Here, it is important to clarify our implementation of testing the global agreement between results of modeling and the IBEX ribbon observations. Assuming the IMSF direction of (ecliptic longitude$=-120^{\circ}$, latitude$=31^{\circ}$), \citet{sch09} first showed a correlation between the measured ribbon ridge and the $\mathbf{B}\cdot \mathbf{R}=0$ locus obtained from simulations, where $\mathbf{B}$ is the draped field vector outside the heliopause and $\mathbf{R}$ denotes the radial vector along the line of sight. As the IBEX ribbon arc was almost circular, its center was found probably near the  direction of the ISMF, an approach that is is not considered here \citep{funs09}. In our simulations \citep{gry11,stru11,br12,rat12}, we are rather finding the simulated ribbon as a loci of the strongest field outside the heliopause at which $\mathbf{B}\cdot \mathbf{R}=0$. For each assumed ISMF strength and orientation, the resulting simulated ribbon is compared directly to the measured IBEX ribbon ridge of enhanced ENA emission observed at $\sim 2.7$\,keV \citep{funs09}. Note that we do not calculate the ENA fluxes because this would require other (kinetic) treatment of neutrals. Even then, our previous results \citep{gry11} are consistent with the ribbons of simulated ENA fluxes by other authors \citep{chal10,heer11}.

\begin{figure}
\centering
\includegraphics[width=9.cm]{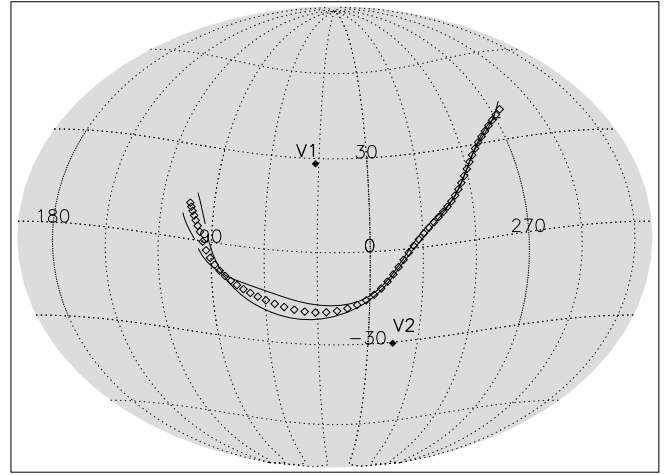}
\caption{The sky-projected positions (galactic coordinates) of computed (solid line) compared to observed (open diamond) IBEX ribbon estimated for the best fit derived in Figure 1 (SW speed $\sim 420$\,km/s). A similar fit is also obtained for V1 TS-crossing time in 2004 (SW speed $\sim 440$\,km/s)}
\end{figure}

\section{Results and discussion}
\subsection{Results for the IBEX He flow}
Following the method described above, we first conducted a sensitivity study to produce a large database of $\sim 2100$ distinct MHD simulations that we compare to the V2 TS crossing on 2007 August 30 and the IBEX ribbon observations. The database is then used to compare galactic latitudes/longitudes for which the MHD model agrees approximately with the measured plasma parameters observed during the V2 TS crossing for different strengths of the ISMF (Fig. 1). We note that the ensemble of solutions is sensitive to the field strength and orientation. Our parametric study shows that our MHD model reproduces the position of the IBEX ribbon and the TS crossing conditions for V2 at $\sim 83.7$ AU, when the ISMF magnitude is in the range of $2.7\pm0.2\, \mu G$ and is pointing away from ecliptic coordinates $(224, 36)\pm3^\circ$, a direction that corresponds to simulation frame angles $\alpha \sim 45^\circ$ and $\beta\sim20^\circ$. This solution was obtained for a slow SW heliographic latitudes range of $\sim\pm 36^{\circ}$ with a density of $\sim 5.2$\,${\rm cm^{-3}}$ and a speed of $\sim 420$\,km/s at 1 AU, consistent with measurements obtained respectively by V2 and by ACE before SW propagation to the TS crossing. The fit to the plasma parameters during the V2 TS crossing is shown in Figure 2 (solid line), and the fit corresponding to the IBEX ribbon is shown in Figure 3. 

The second step in our approach is to determine whether, given the SW conditions that best correspond to the V1 TS crossing on 2004 December 16, and with the ISMF vector derived in step 1, our MHD model provides a good fit to the position of the TS as measured by V1 in the interplanetary magnetic field. As shown in Figure 4 (solid line), a rather good fit is obtained with the same ISMF strength and orientation obtained from V2 and IBEX data analysis but for SW speed $\sim 440$\,km/s and density $5.2$\,${\rm cm^{-3}}$ (SW ram pressure $\sim 1.681$\,nPa). In addition, for the V1 plasma conditions derived above, a good fit (not shown) is also obtained for the IBEX ribbon, confirming the weak sensitivity of the outer heliosheath to SW conditions \citep{izm05a}. 

\begin{figure}
\centering
\includegraphics[width=8.5cm]{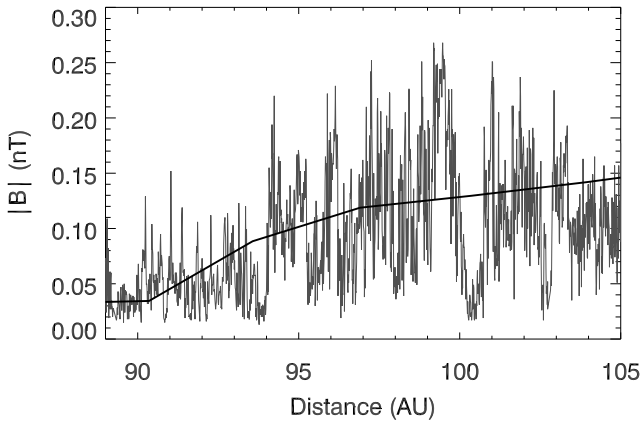}
\caption{The deep heliosphere magnetic field strength. V1 measurements (gray) compared to MHD
model (solid line) for simulation parameters $\alpha=45^\circ$, $\beta=20^\circ$, ISMF strength $2.7\, \mu$G, and SW speed $\sim 440$\,km/s. This model is the best solution obtained in Figure 1 but for the V1 SW conditions (IBEX-based He parameters). }
\end{figure}

\subsection{Need for new constraints}
In the following, we clarify why it is critical to use most of the available {\it in situ} measurements and remote observations to properly constrain the ISMF vector, and more generally to make any diagnostic about the state of the external interstellar medium near our solar system. In the previous section, our analysis was based only on the IBEX ribbon and the V1 \& V2 TS crossings and showed that the solution thus far obtained for the strength of the ISMF depends on the ISM external interstellar conditions assumed. This dependency can be understood from the cumulative effect of several sources of pressure on the heliopause that result from both the gas dynamics and the fields. Therefore, if the given values for the ISM flow are not accurate, any uncertainty in its parameters are immediately propagated to the fields, thereby leading to a degeneracy that hides the exact value of the ISMF. 

As a further example of the relative limitation of any approach that uses only a subset of available observations, we briefly discuss the recent report on the ISM strength as derived from the LISM \lya\ absorption line profile observed for nearby stars when assuming the He LISM parameters obtained from IBEX \citep{zan13}. First, we remark that none of the three models (corresponding respectively to $2$, $3$, and $4$\,$\mu$G) simultaneously fit the \lya\, absorption observed for the four stars thus far selected. In the upwind direction, the best fit is obtained for an ISMF strength of $2$\,$\mu$G, a solution that is rejected by our analysis of the Voyager 1 \& 2 TS crossings (e.g., Figure 1). Although they do not show residuals between model fits and observed \lya\ absorption profiles, one can see that their models with $2$ and $3$\,$\mu$G are closest to the observations (see their Figure 10). This degeneracy calls for additional and distinct observations to constrain the ISMF parameters, including the \lya\ absorption technique that should not be used alone. 

Other constraints are thus required in order to accurately derive the ISMF strength, and more generally the LISM properties. The first suggestion for such a new constraint comes from the SOHO/SWAN experiment that determined the direction from which neutral hydrogen enters the inner heliosheath \citep{lal10}. This direction and He inflow define the HDP, assuming that the deflection of the neutral H flow is caused by an asymmetry in the LISM plasma distribution just outside the HP \citep{lal05}. This suggestion was supported theoretically by MHD-kinetic models that simulated solar wind-LISM interaction ignoring the IMF but including charge exchange between ions and neutrals, with an ISMF vector lying in the HDP \citep{izm05b}. As shown by \citep{pog06} and \citep{pog07}, who included the IMF in the model, the plane defined by LISM velocity and magnetic field vectors does not exactly coincide with the HDP. Building upon this idea, \citet{pog08} used a kinetic model in which the interstellar plasma pressure was not that different from the value used here ($\sim 20$\% smaller in their case), yet they obtained one of the largest asymmetry in the TS shape. For that configuration, \citet{pog08} found that the combined distribution of primary and secondary populations of HI atoms is deflected $\sim 3.8^{\circ}$ in the HDP and $\sim 0.05^{\circ}$ perpendicular to it, confirming that the deflection is produced mainly in the HDP. In contrast, the peaks of distributions of the different HI populations showed slightly larger angles, which allows to bound the angle between the plane formed by the ISMF and He flow vectors and the resulting HDP to be smaller than $\sim 15^\circ$. If this is true, the ISMF vector oriented $\sim 20^{\circ}$ from the corresponding HDP could not be an acceptable solution.

A second suggestion to take off the degeneracy in the ISMF strength and the related ISM properties comes from the long-term monitoring of the interplanetary H\,I flow speed measured upwind from the corresponding strong \lya\ sky background emission line. For reference, we recall that the \lya\ glow backscattered by the interplanetary hydrogen (IPH) was at the heart of the discovery of extrasolar H\,I of interstellar origin near the Earth's orbit. The H\,I speed is derived directly from the Doppler separation between the interplanetary and Earth geocorona \lya\ emission lines. The measured speed corresponds to a specific weighted average HI atoms velocity over the line of sight in the upwind direction. Hydrogen atoms consist of several populations (primary, secondary, etc.) corresponding to the level of charge exchange they suffered. Indeed, interstellar neutral atoms start their charge-exchange with decelerated ions near the heliopause region but are deflected/ionized near the Sun \citep{wal75,bar91}. All kinetic and radiation transfer models thus far proposed for the Lyman-$\alpha$\ line Doppler shift agree that the derived IPH average speed should be smaller than the speed of the flow at infinity \citep{bzo97,sch99,que08,kat11}. The IPH speed also depends on the solar wind activity. It was monitored with SOHO/SWAN and the Space Telescope Imaging Spectrometer onboard the Hubble Space Telescope (HST/STIS) during cycle 22, giving a rather coherent view from both observations and models that the IPH speed is at all times consistent with the  $\sim26$\,km/s measured by Ulysses for the He flow \citep[][Vincent et al., 2013, in prepararation]{vin11}. In contrast, the same monitoring showed that on several dates, the IPH speed does exceed the He velocity of $\sim 23$\,km/s reported by IBEX \citep[see Figure 8 of][]{vin11}. We do not see how the hydrogen population, which suffers important charge-exchanges near the heliopause, could have a line of sight average speed that is comparable to or exceeds the He flow speed \citep{bzo97,que08}. All together, the import of the HDP plane and the independent constraint from the IPH flow speed are clearly in favor of Ulysses He flow velocity of $\sim26$\,km/s.

\subsection{Diagnostic on the LISM bow shock}
The conclusion regarding the LISM He flow parameters that favors the Ulysses measurements as the unique parameters of the LISM bulk flow, also gives us an opportunity to achieve a better insight into the state of the plasma upstream of the heliopause. In the recent context of contradicting conclusions regarding the nature of the LISM bow shock (superfast, subfast, or superslow), we try in the following discussion to clarify the problem using most reasonable parameters of the LISM plasma. Our discussion is based on the fast magnetosonic speed $c_f$:

\begin{equation}
c_f^2 = 0.5\, ( c^2+b^2+\sqrt{(c^2+b^2)^2 -4c^2b^2\cos^2{\theta}} )
\end{equation}

where c and b denotes sound and Alfvenic speed, respectively, and $\theta$ is the angle between the wave vector and the magnetic field vector. The maximum value for the fast speed $c_f^{\ast}$ appears when $\theta=90^{\circ}$ :

\begin{equation}
c_f^{\ast 2} =c^2+b^2
\end{equation}

Having the undisturbed LISM plasma parameters, one can calculate the minimal fast magnetosonic Mach number $M_f = {V /(c_f^{\ast})}$. When the minimal $M_f > 1$, the flow is superfast. In Table 2, the $M_f$ for the LISM parameters in the range $\mathbf{B}_{\mathbf{\infty}} \sim 2.0-4.0\,\mu$G, velocity $V_{\mathbf{\infty}} \sim 22.8-26.4$\,km/s, and $n_{p} \sim 0.05-0.13\, {\rm cm^{-3}}$ assumed at the outer boundary is calculated \citep{sla08}. As one can see for $V_{\mathbf{\infty}} = 26.4$\,km/s, the flow is superfast (or marginally superfast) for the ISMF solution $\mathbf{B}_{\mathbf{\infty}} \sim 2.2\pm0.1\,\mu$G derived in Ben-Jaffel \& Ratkiewicz (2012) and for all possible cases of LISM density \citep{sla08}.
For $V_{\mathbf{\infty}} = 22.8$\,km/s, and the corresponding ISMF values $\mathbf{B}_{\mathbf{\infty}} \sim 2.7\pm0.2\,\mu$G, a solution that is not compatible with SoHO/SWAN and HST observations, the flow is subfast for small $n_p \sim 0.048\, {\rm cm^{-3}}$, but marginally superfast for $n_p >0.095\, {\rm cm^{-3}}$. 

The tabulated results explain that the contradicting conclusions about the nature and existence of the bow shock do not originate from observations but
from different parameters assumed for the LISM flow \citep{br12,zan13,zie13}. Therefore, based on our sensitivity study and the additional arguments that support the Ulysses He flow of 26.4 km/s, the interstellar bow shock does exist for all reasonable LISM proton densities.

\begin{table}
\caption{Minimum fast magnetoacoustic Mach number $M_f$ for different LISM parameters.}
\begin{center}
\scalebox{0.9}{
\begin{tabular}{lcccccr}
\hline
$n_p\,({\rm cm^{-3}})$ $\backslash$ $\mathbf{B}_{\mathbf{\infty}}\,(\mu$G)  &   2.0  &  2.2  &  2.7  &  3.0  &  4.0 & $\mathbf{V}_{\mathbf{\infty}}$\,(km/s) \\
\hline

0.048                                                                                     &  1.10  &  1.03 &  0.88 & 0.81  & 0.63  & 26.4  \\
0.095                                                                                     &  1.36  &  1.29 &  1.13 & 1.05  & 0.84  & 26.4  \\
0.13                                                                                      &  1.47  &  1.40 &  1.25 & 1.17  & 0.96  & 26.4  \\ 
\hline
0.048                                                                                     &  0.96  &  0.89 &  0.76 & 0.70  & 0.54  & 22.8 \\
0.095                                                                                     &  1.18  &  1.12 &  0.99 & 0.91  & 0.73  & 22.8 \\
0.13                                                                                      &  1.28  &  1.22 &  1.09 & 1.02  & 0.83  & 22.8 \\

\hline
\end{tabular} }
\end{center}
\end{table}

\begin{table}
\caption{$\mathbf{B}_{\mathbf{\infty}}$ for Ulysses- and IBEX-based He parameters in ecliptic ($\lambda_{ec}$, $\beta_{ec}$), galactic ($l$,$b$) and simulation ($\alpha$, $\beta$) coordinates.}
\begin{center}
\scalebox{0.8}{
\begin{tabular}{cccccccc}
\hline
$\mathbf{B}_{\mathbf{\infty}}$ & $strength$ & $\lambda_{ec}$ & $\beta_{ec}$ & 
$l$ & $b$ & $\alpha$ & $\beta$ \\ 
& & & & & & & \\
Ulysses & $2.2 \pm 0.1$\,$ \mu$G & $241 \pm 3^\circ$ & $39 \pm 3^\circ$ & $28 \pm 3^\circ$ & $52 \pm 3^\circ$ & $42\pm 2^\circ$ & $0\pm 2^\circ$ \\
IBEX & $ 2.7 \pm 0.2$\,$ \mu$G & $241 \pm 3^\circ$ & $39 \pm 3^\circ$ & $28 \pm 3^\circ$ & $52 \pm 3^\circ$ & $45 \pm 3^\circ$ & $20 \pm 3^\circ$ \\ 
\hline
\end{tabular} }
\end{center}
\end{table}
\section{Summary and conclusions} 
We have shown that the asymmetries observed in the sky background \lya\ emission deep in the heliosphere, in the TS position as measured by V1 and V2, and in the IBEX ribbon of ENAs, can all be explained with a local interstellar magnetic field pointing away from galactic coordinates $(28, 52)\pm 3^\circ $ but with a strength that depends on the LISM He parameters thus far assumed for the external boundary condition of the model. 

For instance, this direction makes an angle of $\sim 42^\circ$ and $\sim 45^\circ$ respectively with the Ulysses and IBEX helium flows. For the Ulysses-based He parameters, the ISMF vector was found in the HDP plane, consistent with an independent analysis of the SOHO/SWAN hydrogen flow deflection angle \citep{heer11,br12}. {However, for the IBEX-based He parameters, the ISMF vector derived here is oriented $\sim 20^{\circ}$ away from the corresponding HDP, in contrast to the solution reported in \citet{mcco12}}. 
Our solution for the ISMF vector indicates degeneracy in the field strength that depends on the assumed external ISM boundary conditions. For the Ulysses He parameters, a rather weak field of $2.2\pm0.1$\,$\mu$G is required while for IBEX-based ISM parameters, a field strength of $2.7\pm0.2$\,$\mu$G is needed. Our sensitivity study showed the need for new constraints (other than only using the IBEX ribbon and the V1 \& V2 TS crossing observations) to avoid that degeneracy. 

A first suggestion comes from the SOHO/SWAN observation of the H\,I direction of deflection. With Ulysses He flow, our solution is in the HDP (uncertainty of $\sim2^{\circ}$), consistent with the general idea that the magnetic field should be in that plane. In contrast, for the IBEX He flow, the solution is $\sim 20^{\circ}$ away from the HDP. A second suggestion to better constrain the ISM He flow comes from the solar cycle monitoring of the interplanetary H\,I flow speed measured upwind from the correspondingly strong Lyman$-\alpha$\,emission line. Independently of the simulations described here, the existing H\,I speeds are consistent with Ulysses He flow of $\sim 26$\,km/s. All elements of the diagnostics reported here seem to support Ulysses He flow parameters for the interstellar state, which allows to diagnostic the nature and existence of the interstellar bow shock. In that frame, our results show that the interstellar bow-shock does exist.

Finally, we want to highlight some critical findings and our recommendations based upon those findings. First, it is possible to make a definite and safe diagnostic about the interstellar medium state in the neighborhood of our solar system when using most existing {\it in situ} and remote observations. The diagnostic provided by the IBEX ENAs spatial distribution is a key ingredient for obtaining the orientation of the ISMF. The simple, pioneering technique to monitor the IPH speed during the solar cycle is also a second, key ingredient to constrain the bulk velocity of the ISM. Measurements made {\it in situ} bring answers to fundamental questions about the physics of the SW-LISM interaction and the local conditions of the time-variable plasma state in the heliosphere. All these ingredients will be much needed in the future if we desire to monitor any change in the interstellar medium conditions that our Sun may encounter on its galactic orbit. These goals will not be achieved without thinking right now about new interstellar probes particularly in the downwind direction, a refurbished version of IBEX, and a far ultraviolet capability to replace the Hubble Space Telescope. Those are the unique capabilities that may provide the key ingredients to allow a reliable diagnostic of the solar system local environment in the future.

\acknowledgements
LBJ acknowledges support from CNES, Universit\'e Pierre et Marie Curie (UPMC) and the Centre National de la Recherche Scientifique (CNRS) in France. RR acknowledges support from the Institute of Aviation. MS acknowledges support by the Polish National Science Center (N N307 0564 40). Authors acknowledge support from HECOLS and ISSI Bern.

\end{document}